\documentclass[jkps,preprint,fleqn,showpacs,showkeys]{revtex4}
\usepackage{graphicx}
\usepackage[dvips]{epsfig}
\usepackage{amssymb}
\usepackage{amsmath}
\usepackage{bm}
\usepackage{mathtools}

\usepackage[labelsep=none,justification=raggedright]{caption}
\usepackage{multirow}
\usepackage{supertabular}
 \usepackage{array} 
 \newcommand\clinewd[1]{\vrule width #1}

\begin{document}
\setcounter{page}{0} 
\title[]{Analysis of $X$ Particle Spectra in Quarkonium Model}
\author{Jeong Hun \surname{Yang}, Su Kyeong \surname{Lee}, Eun-Joo \surname{Kim} }
\author{Jong Bum \surname{Choi}}
\email{jbchoi@jbnu.ac.kr}
\affiliation{Division of Science Education and Institute of Science Education, Chonbuk National University, Jeonju 561-756, Korea}
\date[]{Received 15 June 2015}

\begin{abstract}
Three more $X$ particles are established in the 2014 Particle Data compared
with the 2012 ones. There are now five established $X$ particles named as $X(3872)$,
$X(3900)$, $X(4260)$, $X(4360)$, and $X(4660)$. Since the first $X$ particle $X(3872)$ was discovered
during the search for the remaining P charmonium states, it is valuable to check
whether the established $X$ particles can be explained by quarkonium model. In this paper,
we try to calculate the mass spectra of charmonium system by considering spin-dependent
forces deduced from one gluon exchange diagrams. The confining potential form is taken
to be linear and the free parameters are determined by least squares method comparing
the theoretical and the observed masses of charmonium states.
 
\end{abstract}
\maketitle

\section{INTRODUCTION} \label{sec:s1}
The analysis of charmonium system motivated important steps to understand quantitatively
the mechanism of quark confinement in various hadrons. The first approach was made by
introducing confining potential to estimate the masses of excited states just after the 
discovery of $J/\psi$ \cite{bib-1}. Because the charm quark was taken to be massive, nonrelativistic Schr\"{o}dinger
equation was applied to predict the excited states and the predictions turned out to be quite
successful. Later the same formalism was used to calculate the detailed splittings of spectra
in the more massive system of bottomonium. These successes lead to the consideration of
spin splittings and a systematic derivation was possible with the introduction of one gluon
exchange diagrams \cite{bib-2}.

Although the spin splittings due to spin-spin, spin-orbit, and tensor forces have been
calculated successfully, the form of confining potential has not been determined clearly and
various models were tried to explain quarkonium spectra. The discrimination of different
potential models can be carried out by comparing the spectral pattern of excited states
but the observations of excited states with orbital angular momentum had been tedious
processes analyzing every decaying channels between excited states. During these analyzing processes 
new states with somewhat peculiar aspects of decaying patterns have been
observed and named as $X$ particles \cite{bib-3}.

In order to predict the energy levels of quarkonium states, we need to fix several
parameters such as quark mass, strong coupling constant, and potential parameter. These
parameters are usually taken to be independent from each other, however, they are interrelated
through the exchanges of self-interacting gluons. The dynamical quark mass induced from
the excited quarkonium states includes large gluonic contributions and the strong coupling
constant ${\alpha}_s$ varies according to the value of exchanged gluonic momentum \cite{bib-4}, and the potential
parameter represents just the effects of gluonic interactions. Because these problems are
not resolved as yet from first principles, we have to fix these parameters by considering observed
data and the most explicit data to be considered are the observed energy spectra. The analysis
of $X$ particles in this viewpoint is necessary and we will carry out this analysis in this paper. 

In Sec.\ref{sec:s2}, we will review the observational processes of $X$ particles, and in Sec.\ref{sec:s3}, charmonium
states are calculated with spin-dependent forces. The calculated and the observed $X$ particles are
compared in Sec.\ref{sec:s4} and the final section is devoted to discussions.

\section{OBSERVED $X$ PARTICLES}  \label{sec:s2}
The history of $X$ particles is quite long. For example, in 1974 Meson Table we can 
find five states $X(1430)$, $X(1440)$, $X(1690)$, $X(1975)$, and $X(2500-3600)$. All these states were
unestablished at that time while only 21 meson states were accepted as established for isoscalar
or isovector states. The state $X(1430)$ was renamed as $\eta(1430)$ when the naming scheme
for mesons was changed in 1986. The change of naming scheme had been motivated by former
papers \cite{bib-5} classifying meson states into radially and orbitally excited states of quark-antiquark bound
system. In doing so the spin-dependences of the observed meson spectra played critical roles
and it turned out that the magnitudes of spin splittings could not be easily accounted by
perturbative calculations. A typical example is the mass difference between $\pi(135)$ and $\rho(770)$
which are taken to be the spin singlet and triplet states of isotriplet combination of up
and down quarks. This example clearly shows the problem of definition of quark mass generating
different viewpoints between current quark mass and constituent one. Another example of 
changed name was $X(1440)$ and this state became $\rho(1450)$ in 1988 Particle Data.

In 1988 Particle Data, we have another five states $X(1700)$, $X(1850)$, $X(1935)$, $X(2220)$,
and $X(1900-3600)$ \cite{bib-6}. Of these $X(1850)$ became ${\phi}_{3}(1850)$ and $X(2220)$ was changed into
$f_J$(2220) in 2002 Particle Data just before the discovery of $X(3872)$ in 2003. The
state $X(1700)$ was replaced by $X(1600)$ and $X(1935)$ became $X(2000)$ in 2002 Particle Data \cite{bib-7}
but both states disappeared after the discovery of $X(3872)$. The state $X(1900-3600)$
existed in 1974 as $X(2500-3600)$ which had mass range overlapping with 1S and 1P
states of charmonium system. The narrowness of the widths of these charmonium states prevented
them to be checked as independent states at that time and the extension of the mass
range from 2500 MeV to 1900 MeV could be related to the appearance of charmed particles.
Anyway only two $X$ particles $X(1600)$ and $X(2000)$ remained in 2002 Particle Data, but the
name $X$ was assigned only to $X(3872)$ after its discovery. In 2004 Particle Data, only
one state $X(3872)$ was named as $X$ particle and the meaning of the assignment $X$ had been
changed into the designation for non-quarkonium state \cite{bib-8}.

$X(3872)$ was discovered during the search for the 2P charmonium states and this state
had been established since 2006. In 2006 Particle Data \cite{bib-9}, one component of 2P charmonium
state $\chi_{c2}$ had been tabulated also and it was necessary to check whether the other components
could be detected independently from the observed $X(3872)$. These searches resulted in 
the report of $Y(3940)$ and $Y(4260)$. Two years later, $Y(3940)$ had been changed into
$X(3940)$ and $Y(4260)$ became $X(4260)$ \cite{bib-10}. The state $Y(3940)$ had been found in the $\omega J/\psi$
invariant mass distribution for exclusive B decays into $K\omega J/\psi$ by Belle Collaboration \cite{bib-11}. But
the state is still not established because of lack of confirmations by other experimental groups.
On the other hand, the state $Y(4260)$ was found by BaBar Collaboration in the invariant-mass
spectrum of $\pi^+\pi^- J/\psi$ through the study of initial-state radiation events \cite{bib-12}. This state had been established in the 2008 Particle Data.

Another state reported in 2006 Particle Data is $X(1835)$. This state was observed
by BES Collaboration in the invariant mass spectrum of $p\bar{p}$ pairs from radiative $J/\psi$ decays \cite{bib-13}.
The first suggestion to account for the decay channels was a $p\bar{p}$ baryonium which could
be a tetraquark generator. However, other possibilities such as pseudoscalar glueball
or radial excitation of ${\eta}'$ had been considered also but the state could not be
established because other groups could not confirm the existence of it. Instead of
confirmation of the same state, another state $X(1840)$ had been added to the 2014
Particle Data \cite{bib-14}.

In 2008 Particle Data, two new state $X(3945)$ and $X(4360)$ had been reported
and the state $Y(4260)$ was changed into the established $X(4260)$. The state $X(3945)$
appeared in 2010 Particle Data \cite{bib-15} but disappeared after that. $X(4360)$ stayed as unestablished
state for several years and finally became established one in 2014 Particle Data. Similarly
$X(4660)$ appeared in 2010 Particle Data and became established state in 2014 Particle Data.
In addition there appeared many $X$ particles in 2010 Particle Data such as $X(4050)$, $X(4140)$,
$X(4160)$, $X(4250)$, $X(4350)$, and $X(4430)$. These particles are not established as yet but the
importance to analyze the spectra with respect to a given model is increased. In 2012 Particle
Data \cite{bib-16}, three more $X$ particles $X(3915)$, $X(10610)$, and $X(10650)$ were added and the 
established state $X(3915)$ was assigned to $2P~ \chi_{c0}$ state. Finally, four more states $X(1840)$, $X(3823)$, $X(3900)$,
and $X(4020)$ were added to 2014 Particle Data and these are tabulated in Table 1 according to their appearance in Particle Data.
\renewcommand{\arraystretch}{0.7}
\begin{table}[h]
\footnotesize
{
\caption{$X$ particle states reported in Particle Data.}
\begin{tabular}{|ccccccccc|}
\hline
(years)&         &         &         &                &          &          &          &\\ 
 1974 &  1988   & 2002    & 2004    & 2006           & 2008     & 2010     & 2012     & 2014\\ \hline
    $X(1430)$&         &         &         &                &          &          &          &         \\    
    $X(1440)$&         &         &         &                &          &          &          &         \\     
             &         &$X(1600)$&         &                &          &          &          &         \\    
    $X(1690)$&         &         &         &                &          &          &          &         \\   
             &$X(1700)$&         &         &                &          &          &          &         \\  
             &         &         &         &$X(1835)$       &$X(1835)$ &$X(1835)$ &$X(1835)$ & $X(1835)$\\ 
             &$X(1850)$&         &         &                &          &          &          & $X(1840)$\\  
             &$X(1935)$&         &         &                &          &          &          &          \\    
    $X(1975)$&         &         &         &                &          &          &          &          \\  
             &         &$X(2000)$&         &                &          &          &          &          \\
             &$X(2220)$&         &         &                &          &          &          &          \\ 
$X(2500-3600)$ & $X(1900-3600)$     &         &         &                &          &          &          & \\  
             &         &         &         &                &          &          &          &$X(3823)$\\ 
             &         &         &$X(3872)$&$*X(3872)$      &$*X(3872)$&$*X(3872)$&$*X(3872)$&$*X(3872)$\\     
             &         &         &         &                &          &          &          &$*X(3900)$\\  
             &         &         &         &                &          &          &$*X(3915)$&          \\  
             &         &         &         &$Y(3940)$&$X(3940)$ &$X(3940)$ &$X(3940)$ &$X(3940)$\\  
             &         &         &         &                &$X(3945)$ &$X(3945)$ &          &          \\
             &         &         &         &                &          &          &          &$X(4020)$\\
             &         &         &         &                &          &$X(4050)$ &$X(4050)$ &$X(4050)$\\ 
             &         &         &         &                &          &$X(4140)$ &$X(4140)$ &$X(4140)$\\
             &         &         &         &                &          &$X(4160)$ &$X(4160)$ &$X(4160)$\\ 
             &         &         &         &                &          &$X(4250)$ &$X(4250)$ &$X(4250)$\\ 
             &         &         &         &$Y(4260)$       &$*X(4260)$&$*X(4260)$&$*X(4260)$&$*X(4260)$\\ 
             &         &         &         &                &          &$X(4350)$ &$X(4350)$ &$X(4350)$\\  
             &         &         &         &                &$X(4360)$ &$X(4360)$ &$X(4360)$ &$*X(4360)$\\ 
             &         &         &         &                &          &$X(4430)$ &$X(4430)$ &$X(4430)$\\ 
             &         &         &         &                &          &$X(4660)$ &$X(4660)$ &$*X(4660)$\\  
             &         &         &         &                &          &          &$X(10610)$&$X(10610)$\\ 
             &         &         &         &                &          &          &$X(10650)$&$X(10650)$\\   
\hline
\end{tabular}
~* represents established states.
\label{table:t1}
}
\end{table}
\section{CHARMONIUM STATES WITH SPIN-DEPENDENT FORCES} \label{sec:s3}
The quarkonium system composed of one quark-antiquark pair with masses $m_1$ and $m_2$
can be described by the equation
\begin{equation}
H\psi = E\psi~,
\label{eq:1}
\end{equation}
where the Hamiltonian $H$ can be split as
\begin{equation}
H = H_0~+~\epsilon(r)~+~V_{SD}~
\label{eq:2}
\end{equation}
with
\begin{equation}
H_0 = \sqrt{m_1^2 + \mathbf{p}_1^2}~+~\sqrt{m_2^2 + \mathbf{p}_2^2}~.
\label{eq:3}
\end{equation}
The potential $\epsilon(r)$ represents the spin-independent part and $V_{SD}$ is the spin-dependent
potential. By calculating relativistic propagator corrections in the Wilson loop formed by the
quark-antiquark pair, we obtain \cite{bib-2}
\begin{eqnarray}
V_{SD}(r)&=&\frac{1}{2}\left(\frac{\mathbf{s}_1\cdot \mathbf{L}}{m_1^2}+\frac{\mathbf{s}_2\cdot \mathbf{L}}{m_2^2}\right)
            \left(\frac{1}{r}\frac{d\epsilon(r)}{dr}+\frac{2}{r}\frac{dV_1}{dr}\right)
          +\frac{1}{m_1 m_2}(\mathbf{s}_1 + \mathbf{s}_2 )\cdot \mathbf{L}\frac{1}{r}\frac{dV_2}{dr}  \nonumber\\
          &+&\frac{1}{3m_1 m_2}(3\mathbf{s}_1 \cdot \hat{\mathbf{r}}~\mathbf{s}_2 \cdot \hat{\mathbf{r}} - \mathbf{s}_1\cdot\mathbf{s}_2)V_{3}(r)+\frac{1}{3m_1 m_2}\mathbf{s}_1\cdot\mathbf{s}_2 V_{4}(r) 
\label{eq:4}
\end{eqnarray}
with the potentials $V_i$ defined by
\begin{align}
Te^{-\epsilon(r)T}~\frac{\mathbf{s}_1\cdot\mathbf{L}}{m_1^2}\frac{1}{r}\frac{dV_1 (r)}{dr} = \frac{ig^2}{m_1^2} \int_{0}^{T}dt 
                                     \int_{0}^{T}dt'(t'-t)\mathbf{s}_1\cdot\left\langle \mathbf{B}(\mathbf{x}_1 , t)\mathbf{E}(\mathbf{x}_1 , t')\right\rangle \cdot\nabla(\mathbf{x}_1 , 0)~,         \nonumber\\                                      
Te^{-\epsilon(r)T}~\frac{\mathbf{s}_1\cdot\mathbf{L}}{m_1 m_2}\frac{1}{r}\frac{dV_2 (r)}{dr} = -\frac{ig^2}{m_1 m_2} \int_{0}^{T}dt 
                                     \int_{0}^{T}dt'\mathbf{s}_1\cdot\left\langle \mathbf{B}(\mathbf{x}_1 , t)t'\mathbf{E}(\mathbf{x}_2 , t')\right\rangle \cdot\nabla(\mathbf{x}_2 , T)~,         \nonumber\\                       
Te^{-\epsilon(r)T}~\frac{1}{m_1 m_2}\left[(\mathbf{s}_1\cdot\hat{\mathbf{r}}~\mathbf{s}_2\cdot\hat{\mathbf{r}} - \frac{1}{3}\mathbf{s}_1\cdot\mathbf{s}_2)V_{3}(r) + \frac{1}{3}\mathbf{s}_1\cdot\mathbf{s}_2 V_{4}(r) \right]~~~~~~~~~~~~~~~~~~~~~~~~~~~~~~  \nonumber\\  
=\frac{g^2}{m_1 m_2} \int_{0}^{T}dt\int_{0}^{T}dt'\left\langle \mathbf{s}_1\cdot\mathbf{B}(\mathbf{x}_1 , t)\mathbf{s}_2\cdot\mathbf{B}(\mathbf{x}_2 , t')\right\rangle ~.~~~~~~~~~~~~ 
\label{eq:5}
\end{align}
The expectation values are to be evaluated along the Wilson loop as
\begin{equation}
\left\langle\theta(x)\right\rangle \equiv \int \left[dA^{\mu}\right]Tr\left\{P\left[exp\left(ig\oint_{c}dz_{\mu}A^{\mu}(z)\right)\theta(x)\right]\right\}e^{iS_{YM}} ~~ ,
\label{eq:6}
\end{equation}
where $P$ represents the path-ordered exponential, and the non-Abelian electric and magnetic
fields $\mathbf{E}$ and $\mathbf{B}$ are given by
\begin{eqnarray}
\mathbf{E} & = & -\nabla A^o - \dot{\mathbf{A}} - ig \left[\mathbf{A} , A^o \right]~, \nonumber\\    
\mathbf{B} & = & \nabla\times\mathbf{A} + ig\mathbf{A}\times\mathbf{A}   
\label{eq:7} 
\end{eqnarray}
in terms of the gluonic field $A^{\mu}$. The calculation of $V_{i}$'s can be carried out by
estimating the gluonic correlation functions and because these correlation functions include
nonperturbative contributions we have to introduce some method to parametrize them. One
method to calculate them is to use lattice QCD and there were attempts to estimate
the correlations up to 0.6fm or so \cite{bib-17}. However, in order to check out whether the observed
$X$ particles can be assigned to quarkonium states or not it is better to use explicit
form of confining potential and perturbative approximations to $V_{i}$'s. In this paper we
will take
\begin{equation}
\epsilon(r) = \frac{r}{a^2} - \frac{4}{3}\alpha_{s}\frac{1}{r} + b~~ ,
\label{eq:8}
\end{equation}
and perturbative results are
\begin{eqnarray}
 \frac{1}{r}\frac{dV_{2}(r)}{dr}& = &\frac{4}{3}\alpha_{s}\frac{1}{r^3}~~, \nonumber\\    
 V_{3}(r)& = &\frac{4}{3}\alpha_{s}\frac{3}{r^3}~~,     \\
  V_{4}(r)& = &\frac{4}{3}\alpha_{s}8\pi\delta(\mathbf{r})~~. \nonumber    
\label{eq:9} 
\end{eqnarray}
The form of $V_1(r)$ is fixed by the well-known relation \cite{bib-18}
\begin{equation}
\epsilon(r) = V_2(r) - V_1(r)~~.
\label{eq:10}
\end{equation}
Then the spin-dependent potential becomes
\begin{eqnarray}
V_{SD}(r)& = & \frac{1}{2}\left(\frac{\mathbf{s}_1\cdot\mathbf{L}}{m_{1}^{2}} + \frac{\mathbf{s}_2\cdot\mathbf{L}}{m_{2}^{2}}\right) 
              \left(-\frac{1}{a^{2}r} + \frac{4}{3}\alpha_s\frac{1}{r^3}\right)
               +\frac{4}{3}\alpha_{s}\frac{1}{m_1m_2}\left(\mathbf{s}_1 + \mathbf{s}_2\right)\cdot \mathbf{L}\frac{1}{r^3} \nonumber\\    
         & + &\frac{4}{3}\alpha_{s}\frac{1}{m_1m_2}\left(3\mathbf{s}_1 \cdot \hat{\mathbf{r}}~ \mathbf{s}_2 \cdot \hat{\mathbf{r}} - \mathbf{s}_1\cdot\mathbf{s}_2 \right)\frac{1}{r^3}
               +\frac{2}{3m_1m_2}\mathbf{s}_1\cdot\mathbf{s}_2\frac{4}{3}\alpha_s4\pi\delta(\mathbf{r})~~.  
\label{eq:11} 
\end{eqnarray}
In order to solve the Eq.(1), we have to expand the square root operators in Eq.(3)
by introducing the expansion parameter $M$
\begin{equation}
M =\sqrt{<\mathbf{p}_i^2> + m_i^2} 
\label{eq:12}
\end{equation}
including the momentum expectation value. $M$ is called the effective mass and we get
\begin{equation}
\sqrt{\mathbf{p}_i^2 + m_i^2}\cong\frac{M}{2} + \frac{m_i^2}{2M} + \frac{\mathbf{p}_i^2}{2M}~,
\label{eq:13}
\end{equation}
and the Eq.(1) can be reduced to Schr\"{o}dinger-like equation. But now the singular behaviors
in $\frac{1}{r^3}$ potential and $\delta(\mathbf{r})$ have to be modified in some way and we adopt the slight
change in $\frac{1}{r^3}$ as $\frac{1}{(r+r_q)^3}$ and the $\delta(\mathbf{r})$ is replaced by
\begin{equation}
\delta(\mathbf{r}) = \frac{1}{r_0^3}~e^{-\frac{\pi r^2}{r_0^2}}~.
\label{eq:14}
\end{equation}

For the system of charmonium states, we can take $m_1=m_2$ and the effective masses
are changed from state to state. However, in order to check whether the $X$ particles
can be assigned to charmonium states it is better to reduce the number of parameters,
and so we will take a typical effective mass by comparing observed masses with calculated
masses. The comparison is carried out by least squares method with variation of parameters.
The values of $r_q$ and $r_0$ turn out to be not so much affective to the energy spectra
and therefore we will fix these values as 0.01 and 1.0 GeV$^{-1}$. The parameter $b$ is 
determined by fitting the energy eigenvalue of $1^{3}S_1$ state to the observed mass of $J/\psi$.
Then the remaining three parameters are charm quark mass $m$, potential parameter $a$, and
strong coupling constant $\alpha_s$. These parameters are fully varied to find the best fit
to the observed charmonium spectra. For the fixing of parameter values, we used 10
$(c\bar{c})$ states definitely assigned to each spin-orbit states of charmonium in the 2014 Particle
Data. The states are listed in Table \ref{table:2}
\begin{table}[h]
\caption{10 $(c\bar{c})$ states definitely assigned to each spin-orbit states of charmonium in the 2014 Particle Data.}
\begin{tabular}{|c|c|c|c|}
\hline
\hspace{0.3cm} Name \hspace{0.3cm} & \hspace{0.3cm} $I^{G}(J^{PC})$ \hspace{0.3cm} & \hspace{0.3cm} Mass (MeV) \hspace{0.3cm} & \hspace{0.3cm}Spin-orbit states \hspace{0.3cm}\\ \hline

$\eta_c (1S)$    &  $0^+(0^{-+})$  & 2983.6     &  $1~^{1}S_0$ \\ 
$J/\psi (1S)$    &  $0^-(1^{--})$  & 3096.916   &  $1~^{3}S_1$ \\ 
$\chi_{c0} (1P)$ &  $0^+(0^{++})$  & 3414.75    &  $1~^{3}P_0$ \\
$\chi_{c1} (1P)$ &  $0^+(1^{++})$  & 3510.66    &  $1~^{3}P_1$ \\ 
$h_c (1P)$       &  $?^?(1^{+-})$  & 3525.38    &  $1~^{1}P_1$ \\
$\chi_{c2} (1P)$ &  $0^+(2^{++})$  & 3556.2     &  $1~^{3}P_2$ \\ 
$\eta_c (2S)$    &  $0^+(0^{-+})$  & 3639.4     &  $2~^{1}S_0$ \\
$\psi (2S)$      &  $0^-(1^{--})$  & 3686.109   &  $2~^{3}S_1$ \\ 
$\chi_{c0} (2P)$ &  $0^+(0^{++})$  & 3918.4     &  $2~^{3}P_0$ \\
$\chi_{c2} (2P)$ &  $0^+(2^{++})$  & 3927.2     &  $2~^{3}P_2$ \\
\hline
\end{tabular}
\label{table:2}
\end{table}

The three parameters are determined by finding out the least value of
\begin{equation}
\Delta M = \left[\frac{1}{N}\sum_{i}\left(E_{i}^{cal} - E_{i}^{obs}\right)^2\right]^{\frac{1}{2}}~,
\label{eq:15}
\end{equation}
where $E_{i}^{cal}$ are the calculated masses of charmonium states and $E_{i}^{obs}$ are the observed
masses. $N$ is the number of states used to calculate $\Delta M$ and for the fixing of
parameters we have chosen $N=10$. The variations of $\Delta M$ with respect to each parameter
are shown in Fig.1, Fig.2, and Fig.3. 
The least value of $\Delta M$ is
\begin{equation}
\Delta M = 36.715~ \rm{MeV}
\label{eq:16}
\end{equation}
with the parameter values 
\begin{equation}
m = 4.3~ \rm{GeV},~~~~ a = 2.0~ \rm{GeV}^{-1},~~~~ \alpha_{s} = 0.28~.
\label{eq:17}
\end{equation}
With the determined parameters we can calculate the masses of charmonium states as in Table \ref{table:3} 
%
\begin{figure}[h]
\centering
\includegraphics[width=9cm,height=7cm]{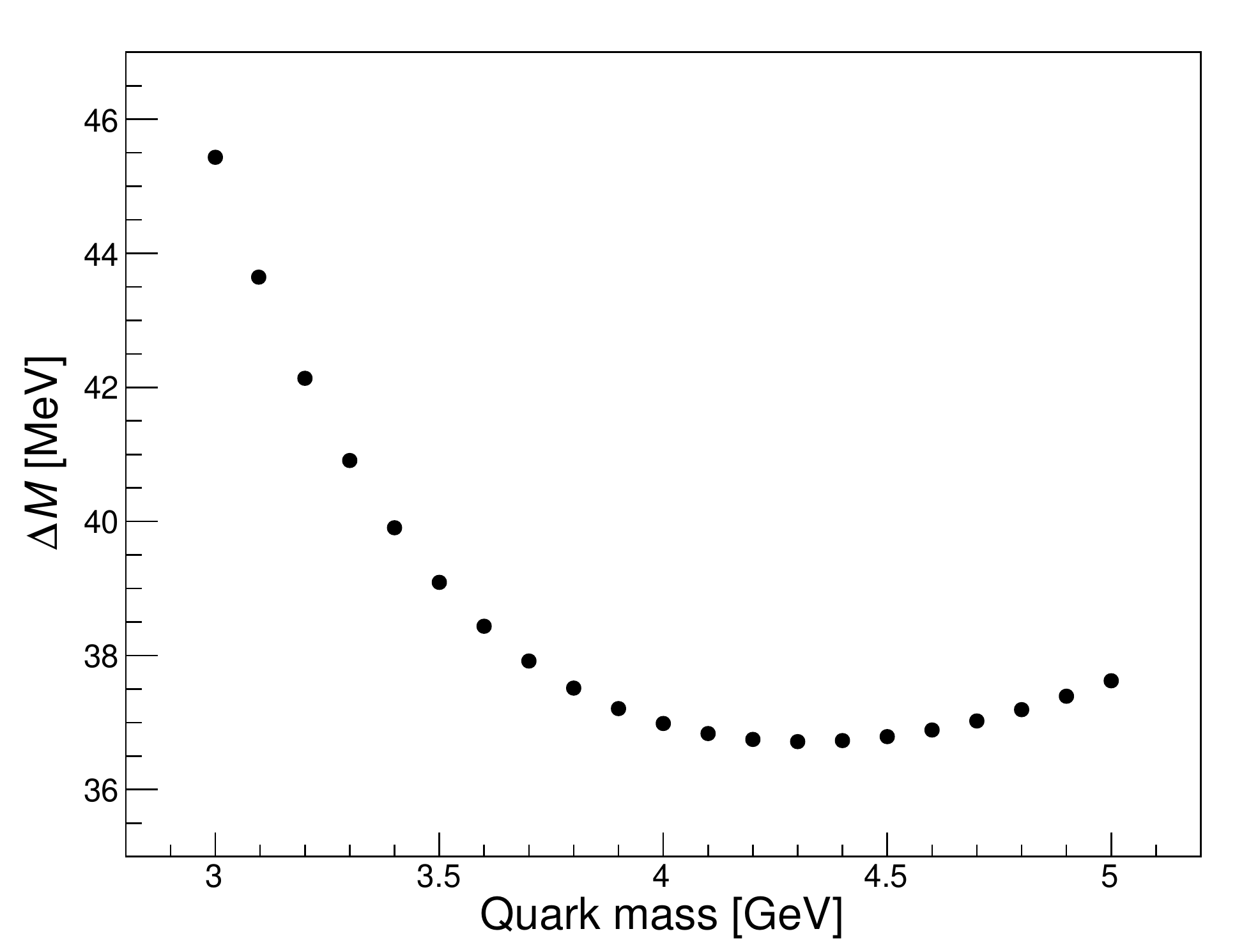}
\caption{The variation of $\Delta M$ with respect to $m$. }
\label{fig:one}
\end{figure}
%
\begin{figure}[h]
\centering
\includegraphics[width=9cm,height=7cm]{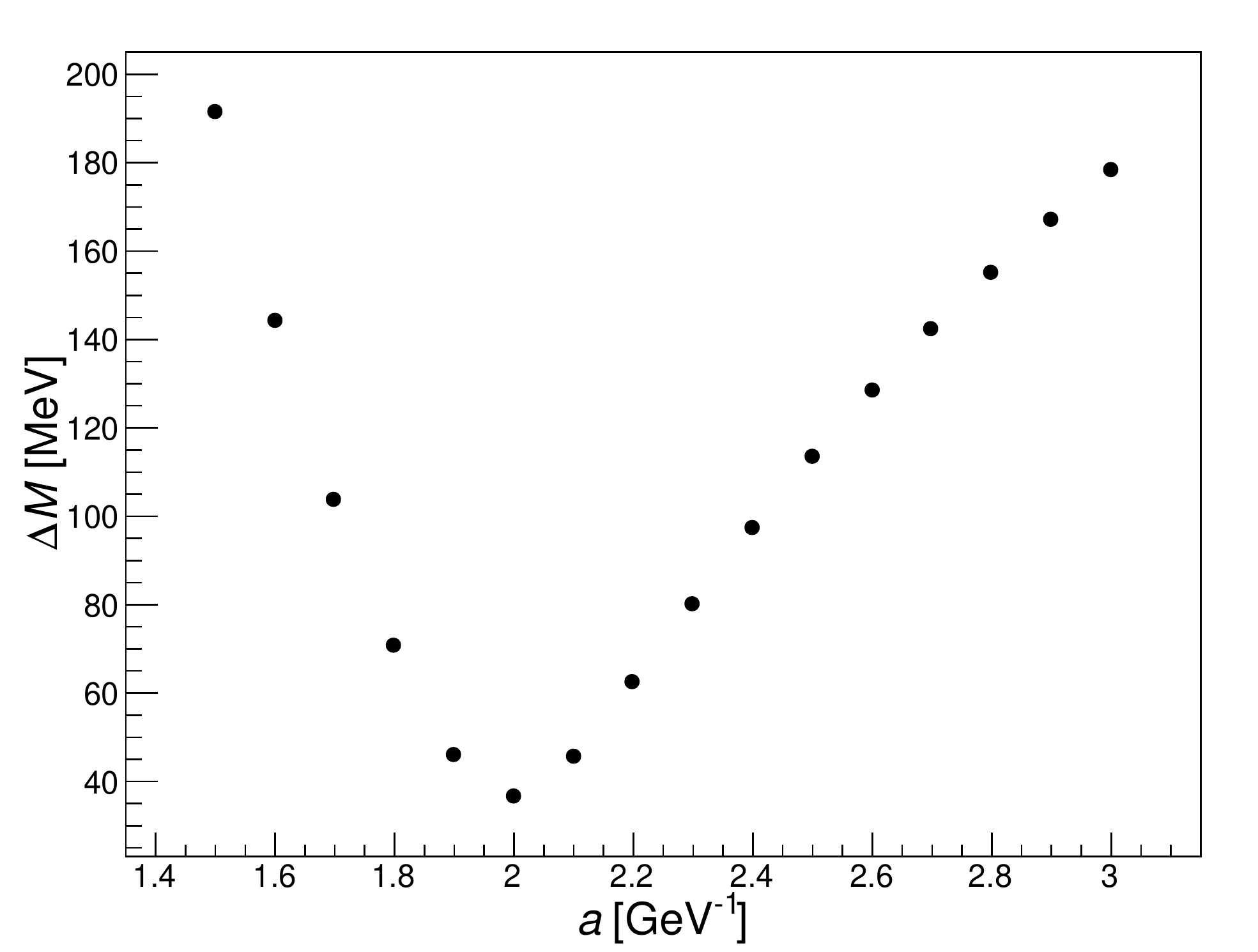}
\caption{The variation of $\Delta M$ with respect to a.   }
\label{fig:one}
\end{figure}
%
\begin{figure}[h]
\centering
\includegraphics[width=9cm,height=7cm]{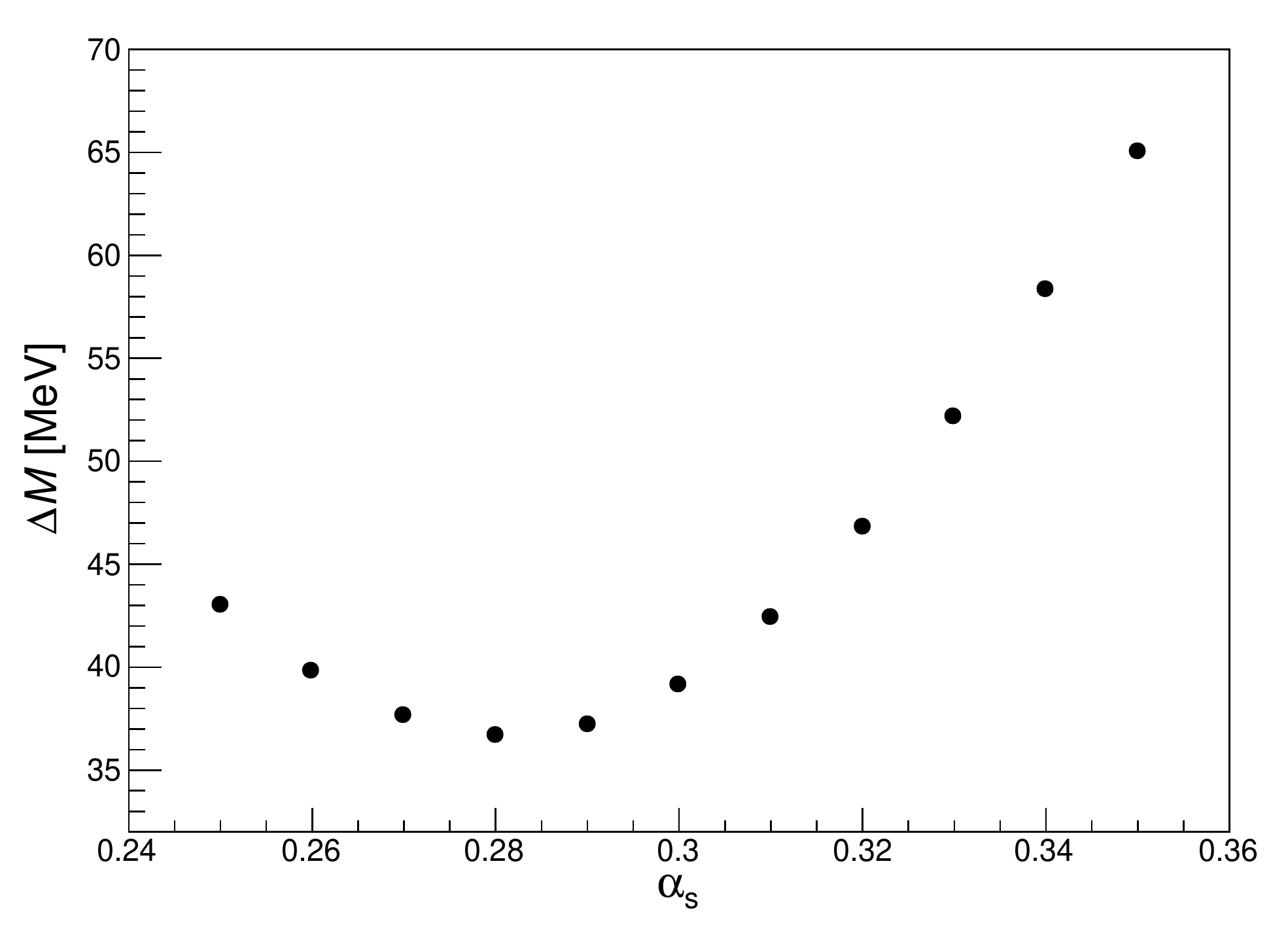}
\caption{The variation of $\Delta M$ with respect to $\alpha_s$. }
\label{fig:one}
\end{figure}
%
\begin{table}[h]
\small
\caption{The caculated masses of $c\bar{c}$ states with $m = 4.3~ \rm{GeV},~ a = 2.0~ \rm{GeV}^{-1},~ \alpha_{s} = 0.28~$. }
\begin{tabular}{|c|c|c||c|c|c||c|c|c||c|c|c|}
\hline
\multirow{2}{*}{n}& \multirow{2}{*}{states} & Energy & \multirow{2}{*}{n}& \multirow{2}{*}{states} & Energy &\multirow{2}{*}{n}& \multirow{2}{*}{states} & Energy & \multirow{2}{*}{n} & \multirow{2}{*}{states}& Energy\\
                  &                         & $(\rm{MeV})$&              &                         &$(\rm{MeV})$    &                  &                         &$(\rm{MeV})$ &                   &                        &$(\rm{MeV})$ \\  \hline
~1~~&~~$^{1}S_0$~~&~~3069.289~~~&~2~~&~~$^{3}P_0$~~&~~3861.784~~~&~1~~&~~$^{3}D_1$~~&~~3758.536~~&~1~~&~~$^{3}F_3$~~&~4013.832~~\\ 
2   &  $^{1}S_0$  &  3658.608   & 3  &  $^{3}P_0$  & 4211.973    & 2  &  $^{3}D_1$  &  4119.056   & 2  &  $^{3}F_3$  &  4341.675 \\ 
3   &  $^{1}S_0$  &  4063.475   & 4  &  $^{3}P_0$  & 4522.449    & 3  &  $^{3}D_1$  &  4436.514   & 3  &  $^{3}F_3$  &  4638.590  \\
4   &  $^{1}S_0$  &  4402.871   & 5  &  $^{3}P_0$  & 4806.507    & 4  &  $^{3}D_1$  &  4725.683   & 1  &  $^{3}F_4$  &  4054.074  \\ 
5   &  $^{1}S_0$  &  4705.338   & 1  &  $^{3}P_1$  & 3502.678    & 1  &  $^{3}D_2$  &  3787.032   & 2  &  $^{3}F_4$  &  4382.869 \\ 
6   &  $^{1}S_0$  &  4983.034   & 2  &  $^{3}P_1$  & 3921.409    & 2  &  $^{3}D_2$  &  4148.184   & 3  &  $^{3}F_4$  &  4680.448 \\ 
1   &  $^{3}S_1$  &  3096.916   & 3  &  $^{3}P_1$  & 4269.803    & 3  &  $^{3}D_2$  &  4466.071   & 1  &  $^{1}G_4$  &  4223.334 \\ 
2   &  $^{3}S_1$  &  3670.321   & 4  &  $^{3}P_1$  & 4578.827    & 4  &  $^{3}D_2$  &  4755.562   & 2  &  $^{1}G_4$  &  4528.210 \\ 
3   &  $^{3}S_1$  &  4071.601   & 5  &  $^{3}P_1$  & 4861.634    & 1  &  $^{3}D_3$  &  3823.186   & 1  &  $^{3}G_3$  &  4175.334 \\ 
4   &  $^{3}S_1$  &  4409.295   & 1  &  $^{3}P_2$  & 3549.774    & 2  &  $^{3}D_3$  &  4185.077   & 2  &  $^{3}G_3$  &  4479.266 \\ 
5   &  $^{3}S_1$  &  4710.737   & 2  &  $^{3}P_2$  & 3966.525    & 3  &  $^{3}D_3$  &  4503.443   & 1  &  $^{3}G_4$  &  4214.091 \\ 
6   &  $^{3}S_1$  &  4987.737   & 3  &  $^{3}P_2$  & 4314.022    & 4  &  $^{3}D_3$  &  4793.273   & 2  &  $^{3}G_4$  &  4518.777 \\ 
1   &  $^{1}P_1$  &  3526.129   & 4  &  $^{3}P_2$  & 4622.503    & 1  &  $^{1}F_3$  &  4023.559   & 1  &  $^{3}G_5$  &  4261.390 \\
2   &  $^{1}P_1$  &  3944.129   & 5  &  $^{3}P_2$  & 4904.928    & 2  &  $^{1}F_3$  &  4351.657   & 2  &  $^{3}G_5$  &  4566.997 \\
3   &  $^{1}P_1$  &  4292.203   & 1  &  $^{1}D_2$  & 3798.540    & 3  &  $^{1}F_3$  &  4648.749   & 1  &  $^{1}G_5$  &  4406.616 \\
4   &  $^{1}P_1$  &  4601.036   & 2  &  $^{1}D_2$  & 4159.998    & 1  &  $^{3}F_2$  &  3981.946   & 1  &  $^{3}G_4$  &  4350.730 \\
5   &  $^{1}P_1$  &  4883.709   & 3  &  $^{1}D_2$  & 4478.081    & 2  &  $^{3}F_2$  &  4309.036   & 1  &  $^{3}G_5$  &  4397.523 \\
1   &  $^{3}P_0$  &  3440.481   & 4  &  $^{1}D_2$  & 4767.712    & 3  &  $^{3}F_2$  &  4605.420   & 1  &  $^{3}G_6$  &  4452.983 \\
 \hline
\end{tabular}
\label{table:3}
\end{table}
\section{COMPARISON OF OBSERVED STATES WITH CALCULATED RESULTS} \label{sec:s4}
There are 9 more established states that can be assigned to charmonium
states if possible. They are listed in Table 4 with possible assignments.
\begin{table}[h!]
\caption{9 states that can be assigned to $c\bar{c}$ states.}
\begin{tabular}{|c|c|c|l|}
\hline
Name             & $I^{G}(J^{PC})$   &~~Mass(MeV)~~   &\hspace{1cm}Possible $(c\bar{c})$ states \hspace{0.5cm}\\ \hline
~~$\psi(3770)$~~~&~~$0^{-}(1^{--})$~~&  3773.15       &\hspace{0.2cm}$2~^3S_1~,~1~^3D_1$ \\
$X(3872)$        &  $0^{+}(1^{++})$  &  3871.69       &\hspace{0.2cm}$2~^3P_1$  \\
$X(3900)^{\pm}$  &  $?(1^{+})$       &  3888.7        &\hspace{0.2cm}$2~^1P_1~,~2~^3P_0,~2~^3P_1,~2~^3P_2$~~~ \\
$\psi(4040)$     &  $0^{-}(1^{--})$  &  4039 $\pm$ 1  &\hspace{0.2cm}$2~^3D_1~,~3~^3S_1$ \\
$\psi(4160)$     &  $0^{-}(1^{--})$  &  4191 $\pm$ 5  &\hspace{0.2cm}$2~^3D_1~,~3~^3S_1$ \\
$X(4260)$        &  $?^{?}(1^{--})$  &  4251 $\pm$ 9  &\hspace{0.2cm}$2~^3D_1~,~3~^3S_1,~3~^3D_1,~4~^3S_1$~~~ \\
$X(4360)$        &  $?^{?}(1^{--})$  &  4361 $\pm$ 13 &\hspace{0.2cm}$3~^3D_1~,~4~^3S_1$ \\
$\psi(4415)$     &  $0^{-}(1^{--})$  &  4421 $\pm$ 4  &\hspace{0.2cm}$3~^3D_1~,~4~^3S_1$ \\
$X(4660)$        &  $?^{?}(1^{--})$  &  4664 $\pm$ 12  &\hspace{0.2cm}$4~^3D_1~,~5~^3S_1$ \\
\hline
\end{tabular}
\label{table:4}
\end{table}
In order to check whether one observed state can be assigned to some charmonium
state or not, it is valuable to calculate $\Delta M$ with inclusion of that state to
the list of fixed states in Table \ref{table:2}~
Of course there should not be overlapping
of states, and therefore $\psi(3770)$ cannot be assigned to $2~^{3}S_1$ state because of
$\psi(2S)$ and $X(3900)^{\pm}$ has to be excluded from $2~^{3}P_0$ and $2~^{3}P_2$ states because of
$\chi_{c0} (2P)$ and $\chi_{c2} (2P)$. Now we can calculate the least square difference $\Delta M$ for
each inclusion of observed state and the results are shown in Table \ref{table:5}~
Since the $\Delta M$ for the 10 fixed states is 36.715 MeV, it seems to be appropriate to take
the upper limit of 50 MeV to discriminate the assignments. Then $\psi(4160)$ cannot be
assigned to $3~^{3}S_1$ state and $X(4260)$ is excluded from any assignment to charmonium
\begin{table}[h!]
\caption{ Calculation of $\Delta M $ of 9 states that be assigned to $c\bar{c}$ states and assignment possibility}
\begin{tabular}{ |c|c|c|c| }
\hline
Name &~~Assigned state~~&~~$\Delta M (\rm{MeV})$~~&~~Possibility~~\\ \hline

~~~$\psi(3770)$~~~~& $1~^{3}D_{1}$ & 35.283 & $\bigcirc$ \\ \hline
$X(3872)$    & $2~^{3}P_{1}$ & 38.082& $\bigcirc$\\ \hline
\multirow{2}{*}{$X(3900)^{\pm}$} & $2~^{1}P_{1}$ & 38.792 &$\bigcirc$ \\ \cline{2-4}
                                 & $2~^{3}P_{1}$ & 36.370 & $\bigcirc$ \\ \hline
\multirow{2}{*}{$\psi(4040)$} & $3~^{3}S_{1}$ & 36.361 &$\bigcirc$ \\ \cline{2-4}
                              & $2~^{3}D_{1}$ & 42.522 & $\bigcirc$ \\ \hline
\multirow{2}{*}{$\psi(4160)$} & $3~^{3}S_{1}$ & 50.214 &$\times$ \\ \cline{2-4}
                              & $2~^{3}D_{1}$ & 41.183 & $\bigcirc$ \\ \hline
\multirow{4}{*}{$X(4260)$} & $3~^{3}S_{1}$ & 64.431 &$\times$ \\ \cline{2-4}
                              & $4~^{3}S_{1}$ & 59.190 &$\times$ \\ \cline{2-4}
                              & $2~^{3}D_{1}$ & 52.992 &$\times$ \\ \cline{2-4}        
                              & $3~^{3}D_{1}$ & 65.986 &$\times$ \\ \hline
\multirow{2}{*}{$X(4360)$} & $4~^{3}S_{1}$ & 37.915 & $\bigcirc$\\ \cline{2-4}
                           & $3~^{3}D_{1}$ & 41.760 & $\bigcirc$ \\ \hline 
\multirow{2}{*}{$X(4415)$} & $4~^{3}S_{1}$ & 35.184 & $\bigcirc$\\ \cline{2-4}
                           & $3~^{3}D_{1}$ & 35.318 & $\bigcirc$ \\ \hline                                                                             
\multirow{2}{*}{$X(4660)$} & $5~^{3}S_{1}$ & 37.737 & $\bigcirc$\\ \cline{2-4}
                           & $4~^{3}D_{1}$ & 39.640 & $\bigcirc$ \\ \hline  
\end{tabular}
\label{table:5}
\end{table}
states. Now except for $X(3872)$ the remaining states have two possible $(c\bar{c})$
assignments and we try to find out the best assignment by considering each combination
of possible assignments. Firstly, the state $X(4660)$ can be assigned to $5~^{3}S_1$ and $4~^{3}D_1$
states and we can compare these assignments by calculating $\Delta M$. Because the state $X(3872)$
can be overlapped on $2~^{3}P_1$ state with $X(3900)^{\pm}$, we will consider these states later and
the first comparison is carried out with only $\psi(3770)$ fixed at $1~^{3}D_1$ state. The
results are shown in the first two rows of Table \ref{table:6}~
It turns out that $X(4660)$
is better suited to $5~^{3}S_1$ than $4~^{3}D_1$, so we can fix $X(4660)$ as $5~^{3}S_1$ $(c\bar{c})$ state.
\begin{table}[h!]
\footnotesize{
\caption{Combination of possible assignments and the calculation of $\Delta M$}
\begin{tabular}{ |c|c|c|c|c|c|c|c|c|c|c| }
\hline
\multirow{2}{*}{~List~} & \multicolumn{9}{c|} {Charmonium states}                                                                                                                         &  ~~$\Delta M$~~ \\ \cline{2-10}
                        & ~$3~^{3}S_{1}$~ & ~$4~^{3}S_{1}$~ & ~$5~^{3}S_{1}$~ & ~$2~^{1}P_{1}$~ & ~$2~^{3}P_{1}$~ & ~$1~^{3}D_{1}$~ & ~$2~^{3}D_{1}$~ & ~$3~^{3}D_{1}$~ & ~$4~^{3}D_{1}$~ & $ (\rm{MeV})$  \\ \hline
 1 &              &              & $X(4660)$ &                 &                 & $\psi(3770)$ &              &              &           & 36.376\\
 2 &              &              &           &                 &                 & $\psi(3770)$ &              &              & $X(4660)$ & 38.187\\ \hline
 3 & $\psi(4040)$ &              & $X(4660)$ &                 &                 & $\psi(3770)$ & $\psi(4160)$ &              &           & 39.746\\
 4 & $\psi(4160)$ &              & $X(4660)$ &                 &                 & $\psi(3770)$ & $\psi(4040)$ &              &           & 51.090\\ \hline 
 5 & $\psi(4040)$ & $\psi(4415)$ & $X(4660)$ &                 &                 & $\psi(3770)$ & $\psi(4160)$ & $X(4360)$ &           & 41.800\\
 6 & $\psi(4040)$ & $\psi(4415)$ & $X(4660)$ &                 & $X(3872)$       & $\psi(3770)$ & $\psi(4160)$ & $X(4360)$ &           & 42.307\\ 
 7 & $\psi(4040)$ & $\psi(4415)$ & $X(4660)$ & $X(3900)^{\pm}$ & $X(3872)$       & $\psi(3770)$ & $\psi(4160)$ & $X(4360)$ &           & 43.141\\
 8 & $\psi(4040)$ & $\psi(4415)$ & $X(4660)$ &                 & ~$X(3900)^{\pm}$ & $\psi(3770)$ & $\psi(4160)$ & $X(4360)$ &           & 41.321\\ \hline 
 9 & $\psi(4040)$ & $X(4360)$    & $X(4660)$ &                 &                 & $\psi(3770)$ & $\psi(4160)$ & $\psi(4415)$ &           & 39.283\\
10 & $\psi(4040)$ & $X(4360)$    & $X(4660)$ &                 & $X(3872)$       & $\psi(3770)$ & $\psi(4160)$ & $\psi(4415)$ &           & 39.972\\ 
11 & $\psi(4040)$ & $X(4360)$    & $X(4660)$ & $X(3900)^{\pm}$ & $X(3872)$       & $\psi(3770)$ & $\psi(4160)$ & $\psi(4415)$ &           & 40.984\\
12 & $\psi(4040)$ & $X(4360)$    & $X(4660)$ &                 & ~$X(3900)^{\pm}$ & $\psi(3770)$ & $\psi(4160)$ & $\psi(4415)$ &           & 38.927\\ \hline
\end{tabular}
\label{table:6}
}
\end{table}
The next comparison is the assignments of $3~^{3}S_1$ and $2~^{3}D_1$ states to $\psi(4040)$
and $\psi(4160)$. These two cases are compared in the third and the forth rows of
Table \ref{table:6}~ 
The case of $3~^{3}S_1$ to  $\psi(4160)$ and $2~^{3}D_1$ to $\psi(4040)$ is excluded because
$\Delta M$ exceeds 50 MeV. Therefore we can now fix $\psi(4040)$ to $3~^{3}S_1$ state and $\psi(4160)$
to $2~^{3}D_1$ state. Another comparison is between the assignments of $4~^{3}S_1$ and $3~^{3}D_1$
states to $X(4360)$ and $\psi(4415)$. These cases are classified according to the remaining
assignments of $2~^{1}P_1$ and $2~^{3}P_1$ states to $X(3900)^{\pm}$ and $X(3872)$. They are shown in the
lower part of Table \ref{table:6}~
It turns out that $X(4360)$ is better suited to $4~^{3}S_1$ state and $\psi(4415)$ is naturally assigned to $3~^{3}D_1$ state.
With these assignments the
best fit to the observed states has $\Delta M = 38.927~\rm{MeV}$ not so different from the
value $\Delta M = 36.715~\rm{MeV}$ with only 10 fixed states. The best fit includes $X(3900)^{\pm}$
as $2~^{3}P_1$ state and therefore $X(3872)$ cannot be included in the spectra of charmonium
states.

We analyzed only five established $X$ particles. Of these five $X$ particles, two
$X$ particle $X(3872)$ and $X(4260)$ turn out to be non-$(c\bar{c})$ states. However, there
exist 18 $X$ particles in 2014 Particle Data and it seems impossible to explain them
with quarkonium model. Other possibilities such as hybrid states or tetraquark
states have to be considered to account for these observed states. In order to
check whether the other $X$ particles can be assigned to charmonium states or not,
we need to repeat the above processes. But for the other unestablished $X$ particles
it may be sufficient to compare with the calculated spectra like in Table \ref{table:3}~
Because we have obtained the best fit to the observed established states, it is necessary
to recalculate the whole spectra with new parameters determined from the 17 states
assigned to charmonium states. The new parameters are
\begin{equation}
m = 5.0~\rm{GeV},~~~~ a = 2.0~\rm{GeV}^{-1},~~~~ \alpha_{s} = 0.28,
\label{eq:18}
\end{equation}
and the calculated results are shown in Table \ref{table:7}
\begin{table}
\footnotesize{
\caption{The calculated results with $m = 5.0~\rm{GeV},~a = 2.0~\rm{GeV}^{-1},~\alpha_{s} = 0.28$.}
\begin{tabular}{ |c|c|c|c|c!{\clinewd{1.5pt}}c|c|c|c|c| }
\hline
\multirow{2}{*}{~States~} & ~Theory~ &\multicolumn{3}{c!{\clinewd{1.5pt}}} {$c\bar{c}$}  &\multirow{2}{*}{~States~} & ~Theory~  &\multicolumn{3}{c|} {$c\bar{c}$} \\ \cline{3-5}\cline{8-10}
  &  (MeV)  & Name    &~Mass(MeV)~& $I^{G}(J^{PC})$ &    &  (MeV)  & Name  &~Mass(MeV)~& $I^{G}(J^{PC})$ \\
\hline
$1~^{1}S_{0}$ &~~3072.704~~&~~$\eta_c (1S)$~~&~~2983.6~~ &~~$0^+(0^{-+})$~~&~$2~^{1}D_{2}$~&~~4144.799~~&~~             ~~&~        ~~&~~             ~~\\  
$1~^{3}S_{1}$ & 3096.916 &  $J/\psi (1S)$  & 3096.916  &  $0^-(1^{--})$  & $2~^{3}D_{1}$ & 4112.396 & $\psi(4160)$    &4191$\pm$5 & $0^-(1^{--})$ \\          
$2~^{1}S_{0}$ & 3655.824 &  $\eta_c (2S)$  & 3639.4    &  $0^+(0^{-+})$  & $2~^{3}D_{2}$ & 4135.587 &                 &           & \\ 
$2~^{3}S_{1}$ & 3665.211 &  $\psi (2S)$    & 3686.109  &  $0^-(1^{--})$  & $2~^{3}D_{3}$ & 4164.523 &                 &           & \\                                
$3~^{1}S_{0}$ & 4047.043 &                 &           &                 & $3~^{1}D_{2}$ & 4449.003 &                 &           & \\                 
$3~^{3}S_{1}$ & 4053.409 &  $\psi (4040)$  &4039$\pm$1 &  $0^-(1^{--})$  & $3~^{3}D_{1}$ & 4416.202 & $\psi(4415)$    & 4421$\pm$4& $0^-(1^{--})$\\                    
$4~^{1}S_{0}$ & 4372.978 &                 &           &                 & $3~^{3}D_{2}$ & 4439.656 &                 &           & \\       
$4~^{3}S_{1}$ & 4377.961 &  $X(4360)$      &4361$\pm$13&  $?^?(1^{--})$  & $3~^{3}D_{3}$ & 4468.898 &                 &           & \\                    
$5~^{1}S_{0}$ & 4662.635 &                 &           &                 & $4~^{1}D_{2}$ & 4725.639 &                 &           & \\                     
$5~^{3}S_{1}$ & 4666.803 &  $X(4660)$      &4664$\pm$12&  $?^?(1^{--})$  & $4~^{3}D_{1}$ & 4692.543 &                 &           & \\                                  
$6~^{1}S_{0}$ & 4928.151 &                 &           &                 & $4~^{3}D_{2}$ & 4716.197 &                 &           & \\                        
$6~^{3}S_{1}$ & 4931.773 &                 &           &                 & $4~^{3}D_{3}$ & 4745.657 &                 &           & \\                        
$1~^{1}P_{1}$ & 3532.609 &  $h_c(1P)$      & 3525.38   &  $?^?(1^{+-})$  & $1~^{1}F_{3}$ & 4015.570 &                 &           & \\                                          
$1~^{3}P_{0}$ & 3459.000 & $\chi_{c0}(1P)$ & 3414.75   &  $0^+(0^{++})$  & $1~^{3}F_{2}$ & 3983.757 &                 &           & \\                                          
$1~^{3}P_{1}$ & 3512.699 & $\chi_{c1}(1P)$ & 3510.66   &  $0^+(0^{++})$  & $1~^{3}F_{3}$ & 4008.207 &                 &           & \\                                                            
$1~^{3}P_{2}$ & 3552.889 & $\chi_{c2}(1P)$ & 3556.2    &  $0^+(2^{++})$  & $1~^{3}F_{4}$ & 4038.824 &                 &           & \\       
$2~^{1}P_{1}$ & 3935.717 &                 &           &                 & $2~^{1}F_{3}$ & 4329.260 &                 &           & \\                    
$2~^{3}P_{0}$ & 3866.116 & $\chi_{c0}(2P)$ & 3918.4    &  $0^+(0^{++})$  & $2~^{3}F_{2}$ & 4296.696 &                 &           & \\                                         
$2~^{3}P_{1}$ & 3916.802 & $X(3990)^{\pm}$ & 3888.7    &                 & $2~^{3}F_{3}$ & 4321.706 &                 &           & \\                                           
$2~^{3}P_{2}$ & 3954.587 & $\chi_{c2}(2P)$ & 3927.2    &  $0^+(2^{++})$  & $2~^{3}F_{4}$ & 4353.035 &                 &           & \\                                         
$3~^{1}P_{1}$ & 4269.587 &                 &           &                 & $3~^{1}F_{3}$ & 4612.958 &                 &           & \\
$3~^{3}P_{0}$ & 4202.176 &                 &           &                 & $3~^{3}F_{2}$ & 4579.867 &                 &           & \\
$3~^{3}P_{1}$ & 4251.117 &                 &           &                 & $3~^{3}F_{3}$ & 4605.269 &                 &           & \\
$3~^{3}P_{2}$ & 4287.814 &                 &           &                 & $3~^{3}F_{4}$ & 4637.096 &                 &           & \\
$4~^{1}P_{1}$ & 4565.092 &                 &           &                 & $1~^{1}G_{4}$ & 4207.537 &                 &           & \\ 
$4~^{3}P_{0}$ & 4499.221 &                 &           &                 & $1~^{3}G_{3}$ & 4171.471 &                 &           & \\
$4~^{3}P_{1}$ & 4546.888 &                 &           &                 & $1~^{3}G_{4}$ & 4200.635 &                 &           & \\ 
$4~^{3}P_{2}$ & 4582.929 &                 &           &                 & $1~^{3}G_{5}$ & 4236.089 &                 &           & \\
$5~^{1}P_{1}$ & 4835.191 &                 &           &                 & $2~^{1}G_{4}$ & 4498.623 &                 &           & \\
$5~^{3}P_{0}$ & 4770.541 &                 &           &                 & $2~^{3}G_{3}$ & 4461.832 &                 &           & \\
$5~^{3}P_{1}$ & 4817.173 &                 &           &                 & $2~^{3}G_{4}$ & 4491.573 &                 &           & \\
$5~^{3}P_{2}$ & 4852.756 &                 &           &                 & $2~^{3}G_{5}$ & 4527.735 &                 &           & \\ 
$1~^{1}D_{2}$ & 3798.327 &                 &           &                 & $1~^{1}H_{5}$ & 4383.196 &                 &           & \\ 
$1~^{3}D_{1}$ & 3766.507 & $\psi(3770)$    & 3773.15   &  $0^-(1^{--})$  & $1~^{3}H_{4}$ & 4341.549 &                 &           & \\
$1~^{3}D_{2}$ & 3789.320 &                 &           &                 & $1~^{3}H_{5}$ & 4376.447 &                 &           & \\
$1~^{3}D_{3}$ & 3817.792 &                 &           &                 & $1~^{3}H_{6}$ & 4417.722 &                 &           & \\ \hline
\end{tabular}
\label{table:7}
}
\end{table}
\section{DISCUSSIONS} \label{sec:s5}
In this paper, we have calculated the energy spectra of charmonium system by
determining the parameters with least squares method. The determined quark mass
turns out to be very large compared wirh the current quark mass deduced from the
electroweak interactions. The difference results from gluonic interactions that increase
with higher excited energies. The effect of large dynamical mass appears as
reduction of spin splittings as they are proportional to the inverse square of quark
mass. Thus the largest difference between the calculated and the observed masses of
$\eta_{c}(1S)$ can be ascribed to the large quark mass determined to fit all the observed
data. Moreover, for S-wave states, the probability of quark pair annihilation is not
negligible and similar considerations lead to the introduction of vacuum condensate idea
to
predict the splittings between the triplet and the singlet states. The second state
with large difference between the calculated and the observed masses is $\psi(4160)$. This
state is assigned to $2~^{3}D_1$ and it seems necessary to check the change of measured
mass from the value around 4160 MeV.

The main conclusion of this paper is that the two established $X$ particle
$X(3872)$ and $X(4260)$ cannot be assigned to charmonium states. As is well known
from the time of discovery, $X(3872)$ is considered to be a state of tetraquark.
However, the calculation of energy states for 4-quark degree of freedom is not
an easy one and the observed mass of $X(3872)$ is not quite different from the
mass of $2~^{3}P_1$ charmonium state. In 2014 Particle Data, $X(3900)^{\pm}$ state is established
and stays nearer to $2~^{3}P_1$ charmonium state than $X(3872)$, but it is possible for
$X(3900)^{\pm}$ to be assigned to $2~^{1}P_1$ state and then $X(3872)$ can be assigned to $2~^{3}P_1$
charmonium state. Further study is needed by considering decay processes and state
mixings to clarify the status of $X(3872)$. In contrast, $X(4260)$ is clearly excluded
from the assignments to charmonium states and therefore it is plausible to check the
calculations of tetraquark states to accommodate $X(4260)$.

Another state to note is $\chi_{c0}(2P)$. This state was named as $X(3915)$ in 2012
Particle Data and established immediately after its discovery. However, the difference
between the observed mass and the calculated mass in Table 3 amounts to 56 MeV.
It may be possible to assign $X(3915)$ to $2~^{3}P_0$ charmonium state but the
significant difference between the calculated and the observed masses could be taken to indicate
other possibilities. Other $X$ particles not discussed in this paper are still unestablished 
and we need to confirm these states with more experiments. Peculiar examples are
$X(1835)$ and $X(1840)$. These states cannot be assigned to any charmonium state and it
is open to question whether they can be accounted as tetraquark state or as even
more complex combination of mesons and baryons. Two heavy $X$ particles $X(10610)$ and
$X(10650)$ are thought to be some states containing bottom quark and we need to analyze
bottomonium system.

In summary, now there exist so many $X$ particles that they cannot be accommodated
as quarkonium sates. Of course some $X$ particles can be explained by quarkonium states
but we need to introduce other possibilities such as tetraquark states, hybrids including
gluonic degrees of freedom, and so on. In order to improve theoretical calculations, we
need to rederive the form of gluon propagator in space-time coordinates to get rid of
the divergence behavior at short distances. More systematic derivations of potential forms
are also long-standing problems in strong interaction phenomena. Further establishments
of more $X$ particles may generate active researches on such subjects resulting in the
quantitative understanding of strongly interacting bound systems.


\begin{references} 
\bibitem{bib-1} E. Eichten {\em et al}., Phys. Rev. Lett. {\bf 34}, 369(1975) ; J. S. Kang and H. J.
                Schnitzer, Phys. Rev. D{\bf 12}, 841(1975) ; {\em ibid}. D{\bf 12}, 2791(1975).
\bibitem{bib-2} E. Eichten and F. Feinberg, Phys. Rev. D{\bf 23}, 2724(1981).
\bibitem{bib-3} S. K. Choi {\em et al}.(Belle Collaboration), Phys. Rev. Lett. {\bf 91}, 262001(2003).
\bibitem{bib-4} N. G. Hyun and J. B. Choi, J. Korean Phys. Soc. {\bf 28}, 20(1995) ; {\em ibid}. {\bf 28}, 671(1995). 
\bibitem{bib-5} J. B. Choi, Phys. Rev. D{\bf 31}, 201(1985) ; S. Godfrey and N. Isgur, Phys. Rev. D{\bf 32}, 189(1985).                
\bibitem{bib-6} G. P. Yost {\em et al}. (Particle Data Group), Phys. Lett. B{\bf 204}, 1(1988).
\bibitem{bib-7} K. Hagiwara {\em et al}.(Particle Data Group), Phys. Rev. D{\bf 66}, 010001(2002).
\bibitem{bib-8} S. Eidelman {\em et al}.(Particle Data Group), Phys. Lett. B{\bf 592}, 1(2004).
\bibitem{bib-9} W. M. Yao {\em et al}.(Particle Data Group), J. Phys. G{\bf 33}, 1(2006).
\bibitem{bib-10} C. Amsler {\em et al}.(Particle Data Group), Phys. Lett. B{\bf 667}, 1(2008). 
\bibitem{bib-11} S. K. Choi {\em et al}.(Belle Collaboration), Phys. Rev. Lett. {\bf 94}, 182002(2005).
\bibitem{bib-12} B. Aubert {\em et al}.(BaBar Collaboration), Phys. Rev. Lett. {\bf 95}, 142001(2005).
\bibitem{bib-13} J. Z. Bai {\em et al}.(BES Collaboraion), Phys. Rev. Lett. {\bf 91}, 022001(2003).
\bibitem{bib-14} K. A. Olive {\em et al}.(Particle Data Group), Chinese Phys. C{\bf 38}, 090001(2014).
\bibitem{bib-15} K. Nakamura {\em et al}.(Particle Data Group), J. Phys. G{\bf 37}, 075021(2010).
\bibitem{bib-16} J. Beringer {\em et al}.(Particle Data Group), Phys. Rev. D{\bf 86}, 010001(2012).
\bibitem{bib-17} Y. Koma and M. Koma, Nucl. Phys. B{\bf 769}, 79(2007).
\bibitem{bib-18} D. Gromes, Z. Phys. C{\bf 22}, 265(1984).

\end{references}
\end{document}